# Machine Learning Approach for Device-Circuit Co-Optimization of Stochastic-Memristive-Device-Based Boltzmann Machine

Tong Wu[1†], Huan Zhao[2‡], Fanxin Liu[3], Jing Guo[1*], Han Wang[2*]

[1]*Department of Electrical and Computer Engineering, University of Florida, Gainesville, FL 32611, USA*

[2]*Ming Hsieh Department of Electrical Engineering, University of Southern California, Los Angeles, CA 90089, USA*

[3]*Collaborative Innovation Center for Information Technology in Biological and Medical Physics, and College of Science, Zhejiang University of Technology, Hangzhou 310023, P. R. China*

[†]the authors contributed equally to this work.

*Correspondence email: han.wang.4@usc.edu (H.W.), guoj@ufl.edu (J. G.)

**A Boltzmann machine whose effective "temperature" can be dynamically "cooled" provides a stochastic neural network realization of simulated annealing, which is an important metaheuristic for solving combinatorial or global optimization problems with broad applications in machine intelligence and operations research. However, the hardware realization of the Boltzmann stochastic element with "cooling" capability has never been achieved within an individual semiconductor device. Here we demonstrate a new memristive device concept based on two-dimensional material heterostructures that enables this critical stochastic element in a Boltzmann machine. The dynamic cooling effect in simulated annealing can be emulated in this multi-terminal memristive device through electrostatic bias with sigmoidal thresholding distributions. We also show that a machine-learning-based method is efficient for device-circuit co-design of the Boltzmann machine based on the stochastic memristor devices in simulated annealing. The experimental demonstrations of the tunable stochastic memristors combined with the machine-learning-based device-circuit co-optimization approach for stochastic-memristor-based neural-network circuits chart a pathway for the efficient hardware realization of stochastic neural networks with applications in a broad range of electronics and computing disciplines.**

The similarities between statistical mechanics and combinatorial optimization fields, as described by Kirkpatrick et al,[1, 2] has stimulated extensive interest and progresses in developing algorithms and methods based on simulated annealing (SA) for solving optimization problems.[2] These methods and algorithms have found extensive applications in a variety of important problems such as computer-aided circuit design,[3] power systems,[4, 5] fingerprint matching,[6] scheduling[7, 8] and routing.[9, 10] Artificial neural networks (ANNs) have been widely adopted to solve a broad range of problems including optimization, voice recognition, computer vision, and electronic design automation.[11-14] A hardware accelerator based on an ANN for the SA algorithms can significantly improve the computational efficiency for solving the combinatorial optimization problems. The Boltzmann machine (BM), whose dynamics is similar to the thermodynamics of a natural physical system, is especially suitable for performing the SA algorithms for optimization.[15] One major challenge, however, is an efficient hardware realization of the stochastic artificial neurons in the BM. To perform the SA tasks, a combinatorial optimization problem can be mapped to an imaginary physical system, whose energy is described by the cost function of the optimization problem. Furthermore, the "temperature" needs to be cooled down in the simulated annealing process, which requires that the effective "temperature" of the stochastic artificial neuron to be dynamically tunable in the BM. A transistor-circuit-based approach is inefficient in implementing a stochastic artificial neuron. Although memristors have been explored for efficient hardware implementation of the ANN structures,[16-23] the following issues need to be addressed. First, the artificial neuron implemented by a memristor needs to be stochastic and shows Boltzmann-like statistics. Second, the effective "temperature" of the stochastic memristor needs to be dynamically tunable. Here we show that by exploiting the unique material properties of the two-dimensional material system, a tunable stochastic memristor following Boltzmann statistics can be realized as the neuronal element in the BM for SA.

Figure 1a shows the schematic of the heterojunction device structure. This vertical memristor is constructed on 285 nm Si/SiO$_2$ (285 nm) substrate with 4 nm oxidized boron nitride, i.e. BNO$_x$, as the resistive switching medium.[24] The top electrode consists of silver while the bottom of the BNO$_x$ layer forms van der Waals interface with multi-layer WSe$_2$. Figure 1b shows a cross-sectional scanning transmission electron microscopy (STEM) image of the memristor and an electron energy loss spectroscopy (EELS) mapping of the material composition, which clearly reveals the device structure consisting of the crystalline layered WSe$_2$ and amorphous BNO$_x$. The silver metal layer serves as an active electrode, which ionizes upon the application of electrical bias and drifts through the BNO$_x$ layer. The silver ions are reduced to silver atoms as they receive electrons from the bottom electrode to form a silver filament inside the BNO$_x$ layer, which switches the device from the high resistance state to the low resistance state. The multi-layer WSe$_2$ under the BNO$_x$ serves not only as an inert electrode connected to the external bias, but also as a gate-tunable series resistor that can vary the potential distributions between the WSe$_2$ and BNO$_x$ layers. The external bias voltage, i.e. $V_{bias}$ as indicated in Figure 1a, includes the potential change across both the WSe$_2$ layer and the BNO$_x$ filamentary medium. Since the resistance of the WSe$_2$ layer can be modulated over six orders of magnitude through electrostatic gating (see supplementary information section S1), the gate bias can hence significantly tune the effective potential drop across the BNO$_x$ layer for a given $V_{bias}$. In a typical device, the WSe$_2$ layer reaches its minimum conductance at -18 V back-gate voltage (see supplementary information section S1). Figure 1c shows several cycles of the typical set-reset hysteresis loops of the memristor under a back-gate voltage of 50 V. The set voltage is around ~0.9 V under this bias condition and stochastically varies over different switching cycles. After the device is switched on, it spontaneously returns to the off-state once V$_{bias}$ becomes smaller than ~ 0.3 V, indicating that the state of the device is volatile. This state volatility is a desirable feature of stochastic sampling devices. Unlike the memristors designed for data storage applications, filament volatility is a preferable characteristic in these devices since faster stochastic sampling and simpler peripheral circuitry can be enabled if the device spontaneously resets to the high resistance state after each sampling event. Such features can be obtained in this device at relatively small current compliance (~50 pA). The low compliance current limits the size of the conductive filament formed during the sampling to reduce its stability. Figure 1d shows the dynamic measurement on the device showing the set and reset time-scale of the device at the gate voltage $V_g = 50$ V. A bias voltage of 1.8 V is applied for 45 ms to set the device, followed by a 0.15 V pulse train to read the memristor state at 7.5 ms time intervals. 10 set-read cycles are shown in Figure 1d. The device always spontaneously self-reset within 7.5 ms, and hence can generate successive sampling without any intentional reset operation.

Figure 2a shows the statistics of the set voltages extracted from 30 set-reset cycles, which follows a Gaussian distribution with a mean value of 0.93 V and a standard derivation of 0.18 V. To understand the stochastic characteristics, a three-dimensional (3D) kinetic Monte Carlo (KMC) method is used to simulate the filament formation and SET process of the memristor device (see supplementary information section S4). The KMC simulation describes the hopping events stochastically by an exponential probability distribution. The agreement between the experiment and simulation indicates that the distribution of the SET voltage is physically due to the stochastic hopping properties of the ions in the filament formation process.[25]

The stochastic ionic movement that dictates the filament formation process in this device provides a platform for realizing exponential class sigmoidal distribution function. Here, we define $P_{SET}(t < t_0, V_{bias})$ as the probability that the device will set within time $t_0$ for a given external voltage bias across the memristive device. Figure 2b shows the experimentally measured distribution of $P_{SET}(t < t_0, V_{bias})$ as a function of the SET voltage shifted with respect to $V_0$ at three different gate biases: -10 V, 20 V, and 50 V, respectively. $V_0$ is the 50% probability bias voltage point, i.e. $P_{SET}(t < t_0, V_{bias} = V_0) = 0.5$. In each test, a voltage $V_{bias}$ is applied between the Ag electrode and $WSe_2$ for $t_0$=300 ms. For each gate bias, this procedure is repeated 50 times at each value of the $V_{bias}$ to obtain the set probability. Measurements at different $V_{bias}$ conditions lead to Figure 2b, which shows the set probability as a function of $V_{bias}$ shifted with respect to $V_0$. As shown in Figure 2b, for $V_{bias}$ significantly lower than $V_0$, the probability of setting the device within a certain time ($t_0 = 300$ ms) approaches zero while for $V_{bias}$ sufficiently higher than $V_0$, this probability is close to unity. In the intermediate region of the probability distribution, a sigmoidal transition region exists where the probability increases as $V_{bias}$ increases. Furthermore, the gate modulated tunable Fermi level and charge density in the $WSe_2$ layer allow the dynamic tuning of the transition region spread in such distribution functions. At voltage biases where the WSe$_2$ layer becomes more resistive (e.g. $V_g$ = -10 V), the effective portion of the potential drops across the memristive switching medium becomes smaller. Hence the bias voltage will be less effective in modifying the probability distribution, resulting in the probability transition occurring within a wider spread of the bias voltage. On the other hand, a WSe$_2$ layer with higher conductance (e.g. $V_g$ = 50 V) as tuned by the gate bias tends to decrease the spread of the sigmoidal transition region of the distribution. Based on the Markovian dynamics approximation, which is valid when the thermal equilibration rate is much larger than the ion hopping rate, the SET probability within $t < t_0$ can be expressed in an exponential form, $P \approx 1 - e^{-\gamma t_0}$, where $\gamma$ is a parameter proportional to the average hopping rate. In the hopping transport regime, the average rate is exponentially sensitive to the applied bias $V_{bias}$, $\gamma \approx \alpha e^{(\beta V_{bias})}$, which results in a double exponential form for the probability, $P(V_{bias}) \approx 1 - e^{-\alpha t_0 e^{(\beta V_{bias})}}$, where $\alpha$ and $\beta$ are constants that are only dependent on the material properties and device structures. This double exponential function asymptotically approaches and can be further simplified to a distribution function that resembles Fermi-Dirac distribution (see supplementary information section S5), which is used to describe the experimentally obtained distribution probability as a function of $V_{bias}$,

$$P(V_{bias}) \approx \frac{1}{1+\exp\left(-\frac{V_{bias}-V_0}{T_V}\right)} = S\left(\frac{V_{bias}-V_0}{T_V}\right) = 1 - f_0\left(\frac{V_{bias}-V_0}{T_V}\right), \quad (1)$$

where $V_0$ is the voltage at which there is 50% probability to set the memristor within 300 ms, and $T_V$ is a characteristic scale in the unit of voltage resembling the temperature effect in Fermi-Dirac distribution, termed as the effective "temperature". Here $S$ is the sigmoid function, sometimes also called the logistic function, and $f_0$ is the Fermi-Dirac distribution function. Based on the values of $V_0$ and $T_V$ extracted from the experimental data and the analytical fit, $V_0$ increases as the WSe$_2$ layer becomes more resistive, which is consistent with our previous discussion. In addition, $T_V$ also increases with increasing resistance in the WSe$_2$ layer. Figure 2b clearly demonstrates the sampling of an exponential class sigmoidal function with the tunable spread of the transition region, reminiscent of the Fermi-Dirac distribution in statistical physics.

To understand the dependence of the effective "temperature" on the applied gate voltage, a behavioral model as shown in the inset of Figure 3a is developed. The device is modeled as a memristor in serial combination with a WSe$_2$ layer modulated by the gate electric field. The gate voltage modulates the resistance of the WSe$_2$ layer as,

$$R_{\text{FET}}(V_{\text{g}}) = \frac{Z}{(V_{\text{g}} - V_{\text{T}})}, \tag{2}$$

where Z is a constant independent of the gate voltage, $V_{\text{g}}$ is the gate voltage, and $V_T$ is the threshold voltage. The voltage on the intrinsic memristor is a fraction of the applied voltage through a voltage divider relation. As a result, the effective "temperature" can be expressed as,

$$T_{\text{V}}(V_g) = T_{V0} \frac{\overline{R_{\text{M}}} + R_{\text{FET}}}{\overline{R_{\text{M}}}} = T_{V0}\left[1 + \frac{Z'}{(V_{\text{g}} - V_{\text{T}})}\right], \tag{3}$$

where $T_{V0}$ is an effective "temperature" constant, $\overline{R_{\text{M}}}$ is the average resistance of the intrinsic memristor, and $Z' = Z/\overline{R_{\text{M}}}$. As shown in Figure 3a, the model describes the modulation of the effective "temperature" by the gate voltage as observed in the experiment. The design of a "cooling" procedure in SA, therefore, can be translated into the design of a series of gate voltage pulses by mapping the effective "temperature" to the gate voltage through the $T_{\text{V}}(V_{\text{g}})$ relation.

The device demonstrated here can enable a compact, single-device implementation of the stochastic artificial neurons in a BM for SA. Figure 3b shows a schematic block diagram of a BM-based circuit, in which the weighted sum can be computed by a standard memristor crossbar array (CBA),[26, 27] and the stochastic artificial neurons can be implemented by the two-dimensional material based memristor devices demonstrated here. In addition, a voltage amplifier is used to amplify the output of the CBA and provides the input voltage to the stochastic artificial neurons. A readout circuit block reads the state of the stochastic artificial neuron device, and provides a binary voltage input to the CBA. The "cooling" procedure in SA can be achieved by designing the applied gate voltage on the stochastic artificial neuron device as described before.

As an example of applying the BM with SA to solve combinatorial optimization problems, a school timetabling problem is solved.[28, 29] The timetabling problem requires assigning resources including teachers (T), classrooms (R), and course subjects (C) to classes of students over a number of periods (P) with a combination of constraints (see supplementary information section S6 for the detailed problem definition). It can be mapped into minimizing the energy (or cost) function of an imaginary physical system, whose dynamics is described by an isomorphic BM. For the school timetabling problem solved here, the timetabling information is expressed in terms of a 4$^{th}$ order tensor whose entry values are represented by the neuron states. The coefficients in the expression of the cost function are mapped to the weights which could be implemented by a cross-bar array. The effects of two sources of stochasticity in the stochastic neurons - the standard deviation of $V_0$, $\gamma = \text{std}(V_0)$ and a finite effective "temperature" $T_V$ - on the BM performance in solving a sample scheduling problem, which involves the assignment of 5 teachers and 5 classrooms to 5 courses over 5 class periods, are examined in Figure 3c and 3d, respectively (also see supplementary information section S6). In Figure 3c, the cost function vs. generation is simulated for different values of γ at zero "temperature" $T_V = 0$, which reduces the BM to a Hopfield network with a stochastic threshold $V_0$, whose randomness is characterized by γ. The results show that with γ close to 0, the network suffers from a problem of trapping in local minima of the cost function. While increasing γ solves the problem of trapping, an excessively large γ perturbs the system away from minimum points.

To study the impact of randomness due to the effective "temperature", Figure 3d assumes γ=0 and performs a SA, which has an exponential form of the cooling procedure as,

$$T_i = T_0[1 - 10^{-\alpha_T i}], \tag{4}$$

where $T_i$ is the temperature at the *i*-th generation, $T_0$ is the initial temperature, and $\alpha_T$ is a unitless exponent factor which determines the cooling rate. A larger positive value of $\alpha_T$ corresponds to a slower cooling rate and a smaller positive value of $\alpha_T$ corresponds to a faster cooling rate. The BM dynamics lowers the total cost function stochastically in the SA process, which is equivalent to the search for an optimized solution stochastically. The main panel of Figure 3d shows the cost function vs. the generation number for several different values of $\alpha_T$, with the cooling procedure shown in the inset. A cooling procedure that is too rapid with a small $\alpha_T = 2$ can lead to trapping in local minima, whereas a cooling procedure that is too slow with a large $\alpha_T = 4$ results in excessive perturbation, both of which miss the global optimization stochastically. A careful design of the cooling procedure parameter $\alpha_\text{T}$, therefore, is essential for the optimum performance of the stochastic neural-network circuit in solving the combinatorial optimization problem.

To design the stochastic-memristor-based neural-network circuit, the stochastic nature at both the device and circuit levels needs to be addressed. First, because the circuit characteristics are stochastic, the design objective is in the form of the expectation. Evaluating a data point of the stochastic hardware in the design space requires averaging a sufficiently large number of samples, which can be computationally expensive. Second, the relation between the design objective function and the design space parameters is unknown and can be non-convex. Third, even with a large number of samples, statistical noise still exists in the dataset. For design optimization of the stochastic neural network, we develop a new method by combining the Markov chain Monte Carlo (MCMC) simulations of the device and circuit with Bayesian optimization (BO), and show that this new MCMC-BO approach is especially suitable and efficient to address the stochastic nature. Previous application of the traditional BO method in electronics has been limited to deterministic CMOS circuitry.[30] Stochastic neuromorphic circuits discussed here, however, have fundamentally different operation principles and require device-circuit co-design of stochastic parameters, which can be addressed by the new MCMC method. The BO can use Gaussian process (GP) as a prior, and the new data points in addition to a small initial dataset can be obtained iteratively as the next "best" guess determined by an acquisition function[31]. The method requires only a small dataset and is accommodative to a general design objective function and statistical noise in the dataset.

The results for the design optimization of the BM circuit for SA by using the MCMC-BO method are shown in Figure 4. The optimization objective function, which is defined as the expectation of the cost or energy of the BM, is obtained by using the sample average approximation.[32] To achieve statistical accuracy, it is computed as the average of 3000 generations after an initial 2000 generations of the burn-in phase, whose samples are discarded, in each MCMC simulation of the BM, and it is further averaged over 100 independent chains whose initial states are random.[33] A multivariable design parameter space is formed by $\gamma$ and $\alpha_T$. In each BO step, a new "best" guess data point in the design space, which is determined by the acquisition function of the BO, is computed, and the hyper-parameters of the GP is learnt. Figure 4a shows the design objective function after 5 initial data points and 25 additional BO steps. The number of BO iteration steps determines the balance between the computational efficiency and accuracy, which can be examined by checking the predictive uncertainty of the GP model in BO as discussed below. As shown by the bottom panel of Figure 4a and the highlighted region in Figure 4c, a region around $3.1 < \alpha_T < 3.4$ and $0 < \gamma < 0.37$, is identified as the near-optimal design region. In addition, we tested several types of acquisition functions using the BO optimization, and it is found that the identified optimum design region is insensitive to the specific choice of the acquisition function (see supplementary information section S9).

To quantify the predictive uncertainty of the GP model used in BO, the inset of Figure 4b shows the predictive uncertainty averaged over the entire design space ($2 < \alpha_T < 4$ and $0 < \gamma < 1.0$ V) vs. the BO iteration step number. The result shows a decrease in the average uncertainty in the first 15 steps, and it remains approximately unchanged subsequently.

The main panel of Figure 4b, which resolves the predictive uncertainty in the design space with 30 data points, indicates that the predictive uncertainty is the smallest near the optimum region. As shown in Figure 4c, the efficiency of the design optimization method benefits from the strategy of sampling mostly in the near optimum region, especially in the later steps of BO. Alternatively, the optimization convergence can be checked heuristically by comparing the objective function from *n* BO steps with that from a larger *m* (>*n*) BO steps, by assessing the relative convergence $e_o(n,m) = |o_{min}^n - o_{min}^m|/o_{min}^m$, where $o_{min}^n$ is the minimum objective after *n* BO steps (see supplementary information section S11 and Figure S8).

To confirm that the MCMC-BO method indeed identifies near optimum designs for the stochastic neural-network circuit, we selected a design in the identified optimum region with experimentally accessible devices and "cooling" schedule parameters, and assessed its performance in comparison to the designs outside this region. The existence of a region instead of a single point allows designing the "cooling" schedule parameter $\alpha_T$ in accordance with a given γ value. For example, for an experimental device with a variation of γ ≈ 0.15 V, it is identified that $\alpha_T \approx 3.31$ falls in the identified region. The cooling schedule starts from an initial effective "temperature" $T_0 = 0.5$ V and "cools" down with the exponential schedule, which falls in the range of the "temperature" accessible by modulating the gate voltage of the experimental device as shown in Figure 3a. Figure 4d compares the statistical distribution of the cost function for this design with those of two other designs outside the optimum region. Not only the average cost of this design reduces compared to two other designs, but also the variance of the probability distribution decreases. As a result, for the optimum design, the probability of the cost to be smaller than $C_0$=5.5, *P(Cost<C$_0$)*>0.95, is significantly larger, whereas for $\alpha_T = 4, \gamma = 0$ and $\alpha_T = 2, \gamma = 0.8$ V, the probabilities are *P(Cost<C$_0$)*<0.03 and *P(Cost<C$_0$)*<0.08, respectively. The design within the optimum design space region shows clear performance advantage in terms of a smaller stochastic cost function.

The results here experimentally demonstrate a tunable stochastic artificial neural device enabled by two-dimensional materials interfaced in a hybrid memristive device structure, which shows Fermi-Dirac-like activation behaviors that resemble statistical thermodynamic behaviors of Fermions. The device can provide a highly efficient, single-device realization of the artificial neurons in a stochastic neural-network realization of Boltzmann machine for solving combinatorial optimization problems with SA algorithms. To optimize the design of the Boltzmann machine, we further explore a machine-learning-based strategy to tackle the stochastic nature of the design problem. It is shown that the MCMC-BO method is especially suitable and highly efficient for device-circuit co-optimization of the stochastic-memristor-based neural-network circuit.

**Author contributions**

H.W. and J.G. conceived the project and supervised the research activities. H.Z. and H.W. developed the device structure for generating dynamically tunable sigmoidal distributions. H.Z. fabricated the devices, conducted the electrical measurements and analyzed the experimental data. T.W. and J.G. developed and implemented the SA algorithm and MCMC-BO method. T.W. performed device simulations. F.L. performed the STEM and chemical characterizations. T.W., H.Z., J.G., and H.W. co-wrote the manuscript with contributions from all co-authors.

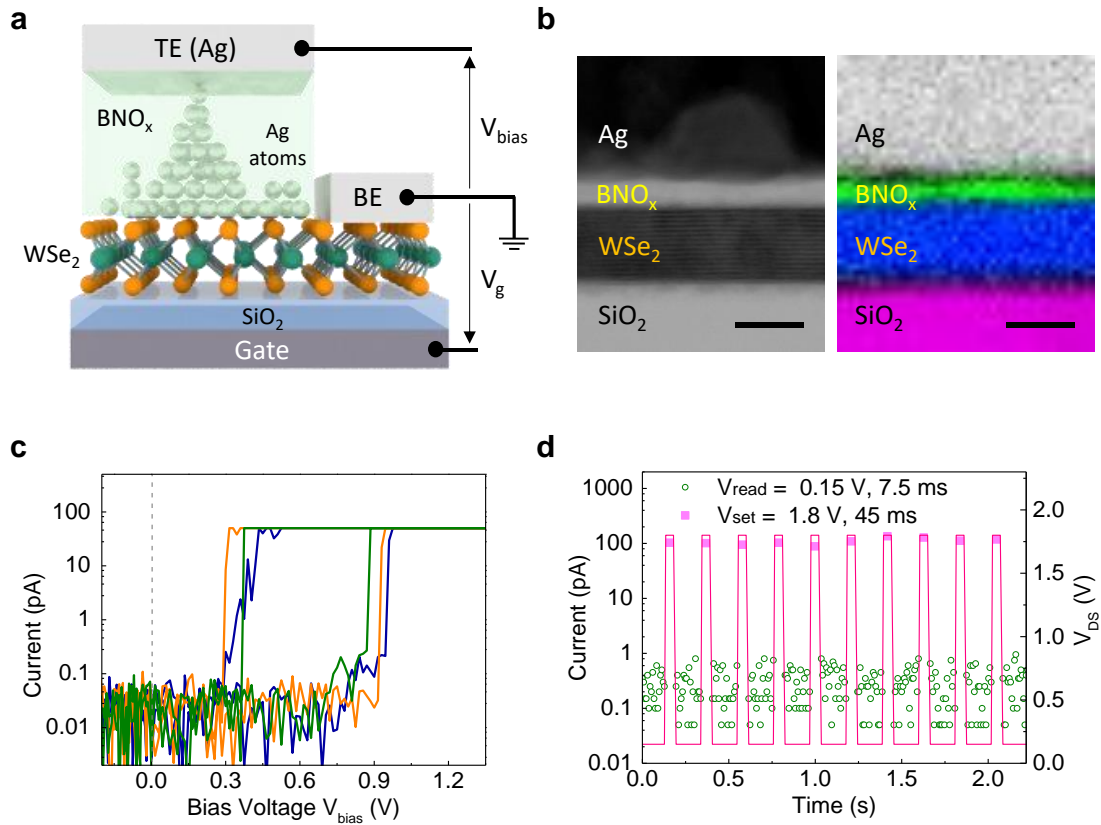

**Figure 1:** (a) Schematic of the hybrid memristive device structure with the $BNO_x$ filament switching layer forming van der Waals interface with the multi-layer $WSe_2$. The top electrode is formed with Ag metal. (b) The cross-sectional STEM image reveals that the device consists of crystalline layered $WSe_2$ and amorphous $BNO_x$. The EELS mapping indicates the material composition in each layer. The scale bars are 10 nm. (c) Three consecutive set-reset switching loops of the device at $V_g = 50$ V. d) The dynamic time domain measurement of the memristor switching at $V_g = 50$ V. In each switching cycle, a 1.8 V, 45 ms voltage pulse was applied between the TE and BE to set the device, followed by a pulse train of 0.15 V amplitude to read the memristor state at 7.5 ms time intervals. Ten set-read cycles were shown. The device always spontaneously self-reset within 7.5 ms. The pink squares are the current level measured during the application of the set pulse. The green circles indicate the current level measured with the read pulse. The set voltage pulse train is shown as the red line with the voltage scale on the right axis.

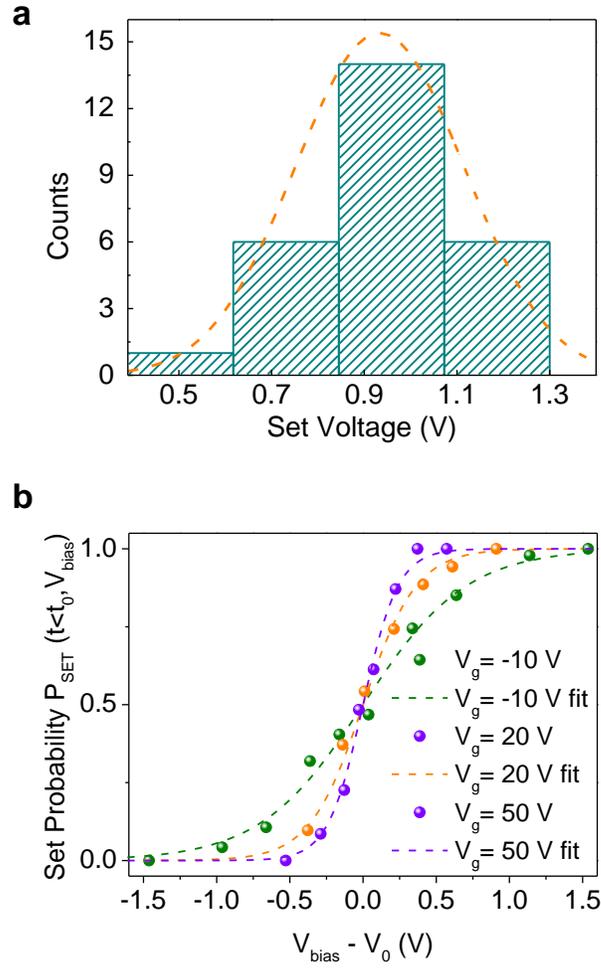

**Figure 2.** (a) The statistics of set voltages extracted from 30 set-reset cycles, under $V_g$ = 50 V. A Gaussian fit of the data is plotted as the orange curve. (b) The probability that the device will set within time $t_0$ = 300 ms as a function of the bias voltage $V_{bias}$ shifted with respect to $V_0$ for $V_g$ = -10 V, 20 V, and 50 V. The experimental data is shown as the dots. The dashed lines show the analytical fit with the sigmoid function eq. (1).

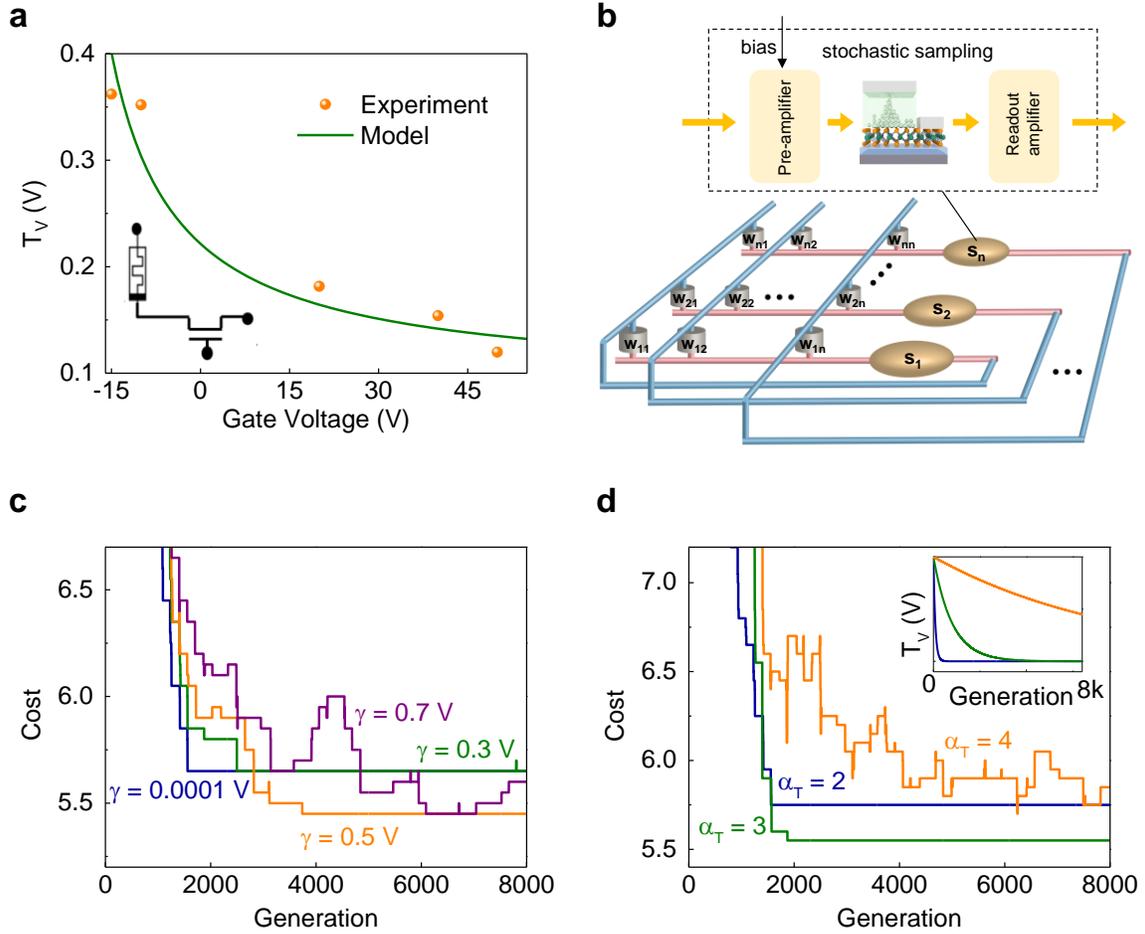

**Figure 3.** (a) Comparison between the model and experiment on the effective "temperature" of the sigmoid distribution as a function of the gate voltage. The inset shows the schematic diagram of the model. (b) Block diagram of a Boltzmann machine designed by using the stochastic memristive neurons for solving a combinatorial optimization problem. (c) The cost vs. generation for different $\gamma$ values, which are the standard deviations in the distribution of $V_0$. The effective "temperature" is 0. (d) The cost vs. generation for different $\alpha_T$ values in SA, whose cooling procedures are shown in the inset, with $\gamma = 0$.

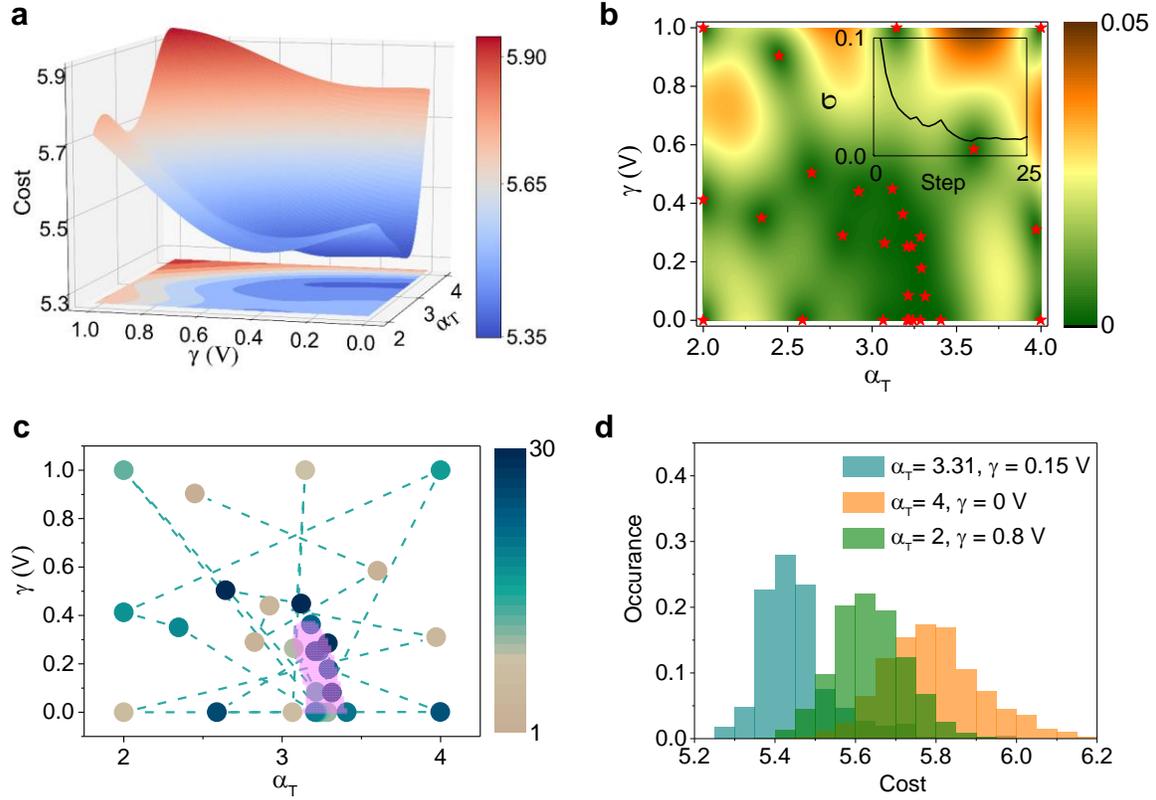

**Figure 4.** Design of the BM for SA by the MCMC-BO simulations with GP. (a) the average cost function vs. $\gamma$ and $\alpha_T$. Here, $\gamma$ is the standard deviation in the distribution of $V_0$, and $\alpha_T$ is the parameter in the cooling procedure. (b) Pseudo color plot of the uncertainty in the average cost as a function of both $\gamma$ and $\alpha_T$. The evaluated design points are shown by the red crosses. The inset shows the average uncertainty vs. the BO step. (a) and (b) are after 5 initial data points and 25 additional BO iteration steps. (c) The trace of the data points for the first 30 data points in BO, and the region with the optimum cost function is highlighted. (d) The stochastic distribution of the cost function for different stochastic design parameters.

# Machine Learning Approach for Device-Circuit Co-Optimization of Stochastic-Memristive-Device-Based Boltzmann Machine

# Supplementary Information

## S1. Electrostatic modulation of current in WSe$_2$

Figure S1 is a plot of back-gate voltage ($V_g$) dependence of the source drain current ($I_{DS}$) of the typical WSe$_2$ field effect device, under $V_{bias}$ = 1 V. The multi-layer WSe$_2$ flake was exfoliated onto 285 nm SiO$_2$ on Si substrate. The contact is formed with 40 nm A$_g$ metal. The $I_{DS}$-$V_g$ curve shows ambipolar behavior with a minimum conductance point at V$_{th}$ = -18 V and an on-off ratio of 10$^6$.

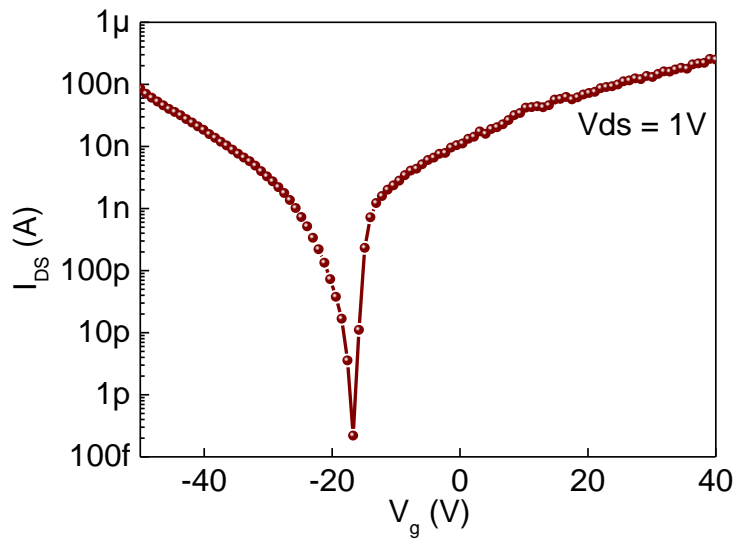

Figure S1. I$_{DS}$-V$_g$ curve showing ambipolar conduction in WSe$_2$ due to field effect modulation.

## S2. Raman spectrum of multilayer WSe$_2$

Figure S2 shows the Raman spectrum of the WSe$_2$ layer in our WSe$_2$/BNO$_x$ memristor. The $E^1_{2g}$ and A$^{1g}$ phonon vibration modes are located at 250 cm$^{-1}$ and 258 cm$^{-1}$, respectively, which is in good agreement with previous study.[S1]

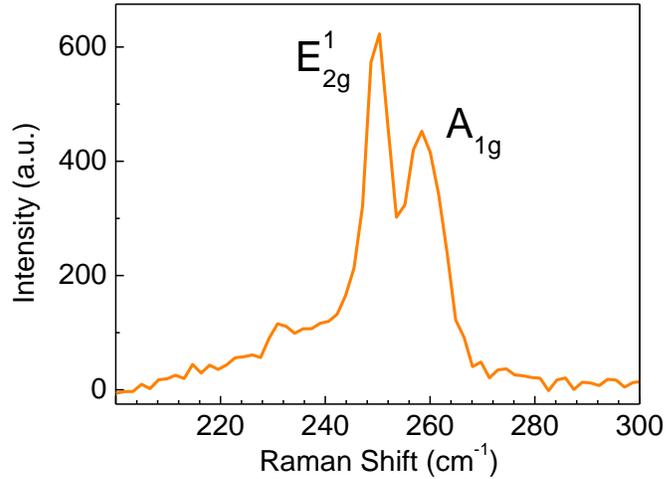

Figure S2. Raman spectrum of multilayer WSe$_2$.

**S3. STEM imaging and EELS mapping**

The interface analysis of the two-dimensional material heterostructures was performed using a transmission electron microscope (Model: FEI Titan Themis G2 with spherical aberration and 4 detectors) combined with an electron energy loss spectroscopy (EELS) gun (Model: Gatan 977). The specimen was first thinned by dual beam focus-ion-beam (FIB) (Model: FEI Helios 450S) under 30 kV acceleration voltage. 5 kV acceleration voltage was then used to polish the interface. It is noted here that the chromium and carbon layers were pre-coated to protect the interface before the FIB thinning process. In the scanning TEM image, a 200 kV acceleration voltage was used. An objective with 41 mm length, 25 mrad convergence angle and 10 mrad receiver angle was used to collect the EELS signals.

## S4. 3D Kinetic Monte Carlo Simulation of the Memristor Device

The kinetic Monte Carlo (KMC) method describes the ion transport stochastically in a three-dimensional (3D) numerical grid, as described in detail in Ref. S2 and outlined below. The 3D KMC simulation of the memristor device starts by calculating the rates of vacancy hopping, generation, and recombination processes. For a device as shown in Figure 1a, the total electric field, $\vec{\varepsilon}$, can be computed from the potential calculated using the current continuity equation. Based on the electric field equations, the rate of ionization, hopping and reduction can be described as $r_{s,e} = r_0 \exp\left(\frac{E_b - d\varepsilon}{k_B T}\right)$, where $r_0$ is a rate constant, $E_b$ is the barrier height, $d$ is the effective hopping distance, $\varepsilon$ is the electric field along the hopping direction, and $k_B T$ is the thermal energy. After the rates are calculated, a random number is generated to determine the event time and another random number is used to determine the type of event. Iterating the above procedures, the evolvement of the filament morphology can be obtained.

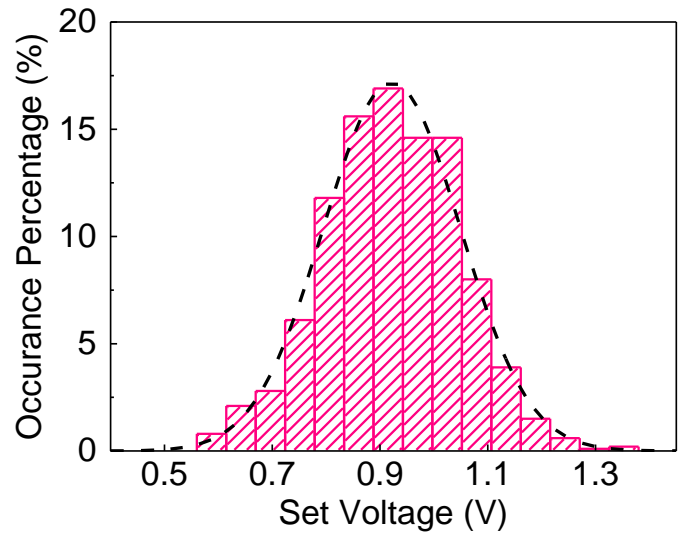

Figure S3. Distribution of $V_{SET}$ obtained by 3D Kinetic Monte Carlo simulation compared to experiment. The black dashed line shows the Gaussian fit to the simulated data.

The set probability $P_{\text{SET}}(t < t_0, V_{\text{bias}})$ can be obtained from multiple KMC simulation samples, as the number of samples which set within $t < t_0$ divided by the total number of KMC simulation samples, which is 1000 in this work. Figure S3 shows the simulated stochastic distribution of the SET voltage, which agrees with the experimental results.

**S5. Asymptotic form of the SET probability**

As described in the main text, the probability function of the stochastic device characteristics can be described by,

$$P_{\text{SET}}(t < t_0, V_{bias}) = 1 - e^{-\alpha t_0 \exp(\beta V_{bias})}. \tag{1}$$

The exponential form as described by eq. (1) has the following limits $P_{\text{SET}}(V_g \to -\infty) = 0$ and $(V_g \to +\infty) = 1$. For a sufficiently large $V_g$ satisfying $\beta V_{bias} > \ln(\frac{10}{\alpha t_0})$, Eq. (1) can be simplified as

$$P_{\text{SET}} \approx 1 - e^{-\alpha t_0 \exp(\beta V_{bias})} \approx \frac{1}{1 + e^{-\alpha t_0 \exp(\beta V_{bias})}}. \tag{2}$$

For a sufficiently small $V_{bias}$, $\beta V_{bias} < \ln(\frac{0.1}{\alpha t_0})$, by using the first order Taylor expansion,

$$P_{\text{SET}} \approx \alpha t_0 \exp(\beta V_{bias}) \approx \frac{1}{1 + \alpha t_0 \exp(-\beta V_{bias})} = \frac{1}{1 + \exp\left(-\frac{V_{bias} - V_0}{T_V}\right)} \tag{3}$$

where $T_V = 1/\beta$ and $V_0 = \ln(\alpha t_0)/\beta$. It indicates that the function asymptotically approaches the sigmoid function above, or equivalent, the Fermi-Dirac distribution function.

**S6. Design of the Boltzmann machine and solution of the timetabling problem**

**1. Boltzmann Machine**

The Boltzmann machine (BM) mimics the thermodynamics of an imaginary physical system, which is used for implementing the simulated annealing algorithm to solve a combinational optimization problem in this work. The network is fully connected with the weights $w_{ki}$ satisfying $w_{ki} = w_{ik}$. The output of each artificial neuron will feedback into the inputs of all other neurons. The energy difference due to the flipping of the $k$-th artificial neuron is,

$$\Delta E_k = \sum_i^N w_{ki} s_i - \theta_k, \qquad (4)$$

and the probability of the state 1 can be obtained according to the probabilistic update rule similar to the Fermi-Dirac distribution function,

$$p_k = \frac{1}{1+exp(-\frac{\Delta E_k}{T})}. \qquad (5)$$

**2. Timetabling problem solved with Boltzmann machine**

The school timetabling problem involves assigning teachers to each class of students. Each class of students will be taught a number of different courses over a number of class periods. Each teacher will require a classroom to teach a particular class of students. In this problem, the class of students is denoted by the classroom that is assigned to them. Let C, T, R, P be the set of courses, teachers, classrooms, and periods. The set of constraints can be described as:

(1). A teacher can only teach $N$ periods of a course in the duration of time table.

(2). During the same period, each course can be given in only one classroom by only one teacher.

(3). During the same period, each teacher can teach only one course in only one classroom.

(4). During the same period, each classroom can be used by only one teacher for only one course.

(5). Each course has a certain teacher.

Suppose a Boolean variable is defined as

$$v_{ijkl} = \begin{cases} 1, & \text{if course i and teacher j are assigned to room k in period l} \\ 0, & \text{otherwise} \end{cases}. \quad (6)$$

To solve this problem, a 4-dimensional BM is assigned. Thus, the energy function can be given as,

$$E = -\frac{1}{2}\sum_{i=1}^{C}\sum_{j=1}^{T}\sum_{k=1}^{R}\sum_{l=1}^{P}\sum_{i'=1}^{C}\sum_{j'=1}^{T}\sum_{k'=1}^{R}\sum_{l'=1}^{P} W_{ijkl,i'j'k'l'} v_{ijkl} v_{i'j'k'l'} - \sum_{i=1}^{C}\sum_{j=1}^{T}\sum_{k=1}^{R}\sum_{l=1}^{P} I_{ijkl} v_{ijkl} \quad (7)$$

In order to solve an optimization problem with BM, we first need to express the energy and the constraints in a single function. Considering the constraints and the conditions, the energy function can be constructed as,

$$E = \frac{C_1}{2}\sum_{i=1}^{C}\sum_{j=1}^{T}\sum_{k=1}^{R}(\sum_{l=1}^{P} v_{ijkl} - N_{ij})^2 + \frac{C_2}{2}\sum_{i=1}^{C}\sum_{j=1}^{T}\sum_{k=1}^{R}\sum_{l=1}^{P}\sum_{\substack{j'=1\\j'\neq j}}^{T}\sum_{\substack{k'=1\\k'\neq k}}^{R} v_{ijkl}v_{ij'k'l}$$

$$+ \frac{C_3}{2}\sum_{i=1}^{C}\sum_{j=1}^{T}\sum_{k=1}^{R}\sum_{l=1}^{P}\sum_{\substack{i'=1\\i'\neq i}}^{C}\sum_{\substack{k'=1\\k'\neq k}}^{R} v_{ijkl}v_{i'jk'l}$$

$$+ \frac{C_4}{2}\sum_{i=1}^{C}\sum_{j=1}^{T}\sum_{k=1}^{R}\sum_{l=1}^{P}\sum_{\substack{i'=1\\i'\neq i}}^{C}\sum_{\substack{j'=1\\j'\neq j}}^{T} v_{ijkl}v_{i'j'kl} + \frac{C_5}{2}\sum_{i=1}^{C}\sum_{j=1}^{T}\sum_{\substack{i'=1\\i'\neq i}}^{T} v_{ij}v_{ij'}$$

$$+ \frac{C_6}{2}\sum_{i=1}^{C}\sum_{j=1}^{T}\sum_{k=1}^{R}\sum_{l=1}^{P}\sum_{\substack{i'=1\\i'\neq i}}^{C} v_{ijkl}v_{i'jkl} + \frac{C_7}{2}\sum_{i=1}^{C}\sum_{j=1}^{T}\sum_{k=1}^{R}\sum_{l=1}^{P}\sum_{\substack{i'=1\\i'\neq i}}^{T} v_{ijkl}v_{ij'kl} \quad (8)$$

where $N_{ij}$ is the limit of the number that teacher $j$ teach course $i$ in the duration of the timetable. The terms with coefficients $C_m$ ($m = 1, 2, \ldots, 5$) each encourage satisfaction of constraints (1)-(5) by counting pairwise products, and the additional terms $C_6$, and $C_7$ are insufficient and necessary conditions of $C_3$, and $C_4$, respectively, strengthening the same constrains. The terms are zero only if the timetable satisfies the corresponding constraints. Together with eq. (7), we can get the expressions of the weight matrix and the additional bias. The weight matrix of the BM network can be expressed as:

$$\begin{aligned}W_{ijkl,i'j'k'l'} =\ & -C_1\delta_{ii'}\delta_{jj'}\delta_{kk'} - C_2\delta_{ii'}(1-\delta_{jj'})(1-\delta_{kk'})\delta_{ll'}\\ & - C_3(1-\delta_{ii'})\delta_{jj'}(1-\delta_{kk'})\delta_{ll'} - C_4(1-\delta_{ii'})(1-\delta_{jj'})\delta_{kk'}\delta_{ll'}\\ & - C_5\delta_{ii'}(1-\delta_{jj'})\delta_{kk'}\delta_{ll'} - C_6(1-\delta_{ii'})\delta_{jj'}\delta_{kk'}\delta_{ll'} - C_7\delta_{ii'}(1-\delta_{jj'})\end{aligned} \quad (9)$$

where $\delta_{mn} = 1$ if $m = n$ and $0$ otherwise. Similarly, each term in this formula corresponds to the constraint in order, and the additional bias can be set as $I_{ijkl} = C_1 \times N_{ij}$. The additional $N_{ij}^2$ term can be ignored since that it only affects the absolute value of energy but does not affect the relative value.

As an example, the problem is set as 5 courses, 5 teachers, 5 rooms, and 5 periods (C=T=R=P=5) for simplicity. The parameters are set as $C_1 = C_2 = C_3 = C_4 = C_5 = C_6 = C_7 = 0.1$. One typical optimization result is shown in Table S1.

**Table S1.** A time table for a day where each cell represents the teachers and their respective courses (teacher: course). The teachers are Alice, Bob, Cindy, David, and Edward. Course codes S, P, H, M, L represents social, physics, history, math, and literature, respectively.

| Class<br>Period | C1 | C2 | C3 | C4 | C5 |
| --- | --- | --- | --- | --- | --- |
| 1 | Alice: S | David: P | Bob: H | Cindy: M | Edward: L |
| 2 | Bob: H | Edward: L | Cindy: M | David: P | Alice: S |
| 3 | Cindy: M | Bob: H | Edward: L | Alice: S | David: P |
| 4 | David: P | Cindy: M | Alice: S | Edward: L | Bob: H |
| 5 | Edward: L | Alice: S | David: P | Bob: H | Cindy: M |

### S7. Gaussian process model used in BO

A Gaussian process model is used as the prior in the Bayesian optimization. The covariance functions of the Gaussian process can be in the form of either the automatic relevance determination (ARD) squared exponential kernel or the ARD Matérn 5/2 kernel.[S3] There are $D+3$ hyperparameters that need to be determined for the Gaussian process: $\theta_{1:D}$, the covariance amplitude $\theta_0$, the observation noise $\nu$, and a constant mean $m$, where $D$ is the dimensionality of the design space. The hyperparameters are determined by optimizing the marginal likelihood of the dataset under the GP.[S3] In the simulations, the hyperparameters are obtained from the dataset by maximizing the marginal likelihood, and the acquisition function is subsequently computed from the mean and covariance of the Gaussian process function. The next data point of inquiry is determined by maximizing the acquisition function.

## S8. Design in the "temperature" parameter space

The design in the standard deviation of $V_0$ and the temperature ($\gamma$, $T_V$) design space is also simulated, as shown in Figure S4. Figure S4a shows the design objective function after 5 initial data points and 25 additional BO steps. As shown by the highlighted region in Figure S4c, a region of $0 < T_V < 0.55$ and $0 < \gamma < 0.32$, is identified as the near-optimal design region. The inset of Figure 4b shows the predictive uncertainty averaged over the entire design space ($0 < T_V < 1.2$ and $0 < \gamma < 1.0$ V) vs. the BO iteration step number. The uncertainty has a decreasing trend during the 25 BO steps. However, it is slightly larger than that in Figure 4b of the main text, which is caused by the large search space including the region where the total cost is too high to be searched further. (That is also why the search space shown in Figure S4a is smaller.) The smallest predictive uncertainty is still near the optimum region.

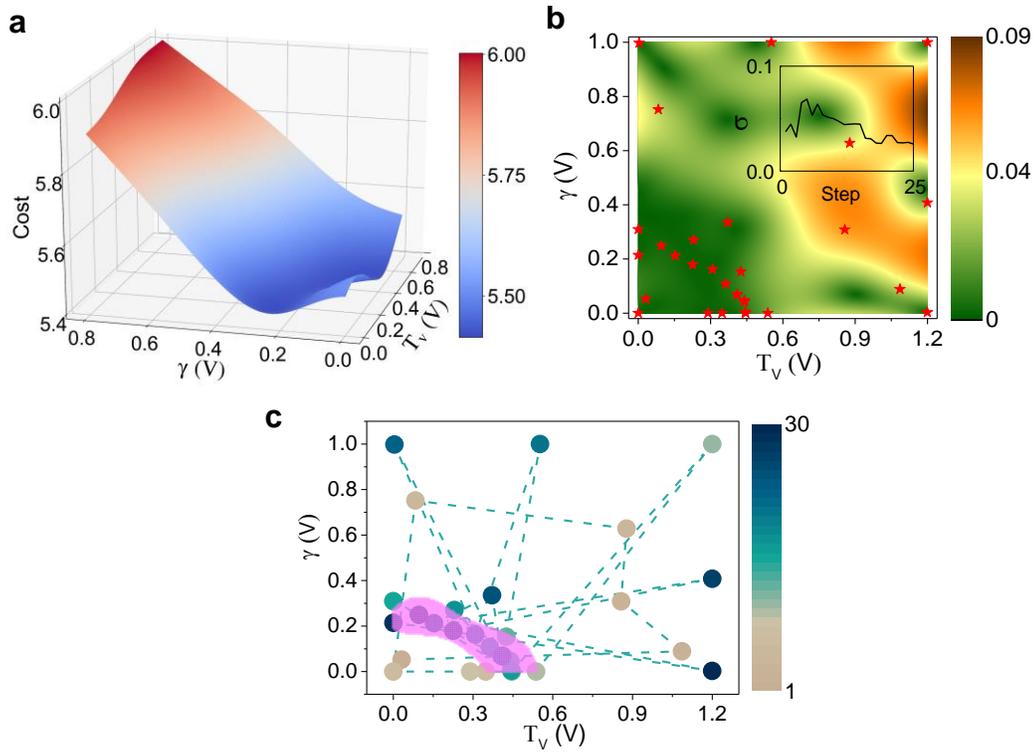

Figure S4. Simulation result of the MCMC-BO search to optimize the design in the ($\gamma$, $T_V$) design space. (a) the average cost function vs. $\gamma$ and $T_V$. (b) Pseudo color plot of the uncertainty. The inset shows the average uncertainty vs. the BO steps. (c) Search step and optimal region.

## S9. Effect of the acquisition function and its parameters

The acquisition function is computed by one of the following three methods: the **probability of improvement** (PI), which maximizes the probability of improving over the best current value; the **expected improvement** (EI), which maximize the expected improvement over the current best; the **upper confidence bound** (UCB) method, which minimizes regret over the course of their optimization by exploiting lower confidence bounds (upper when considering maximization).

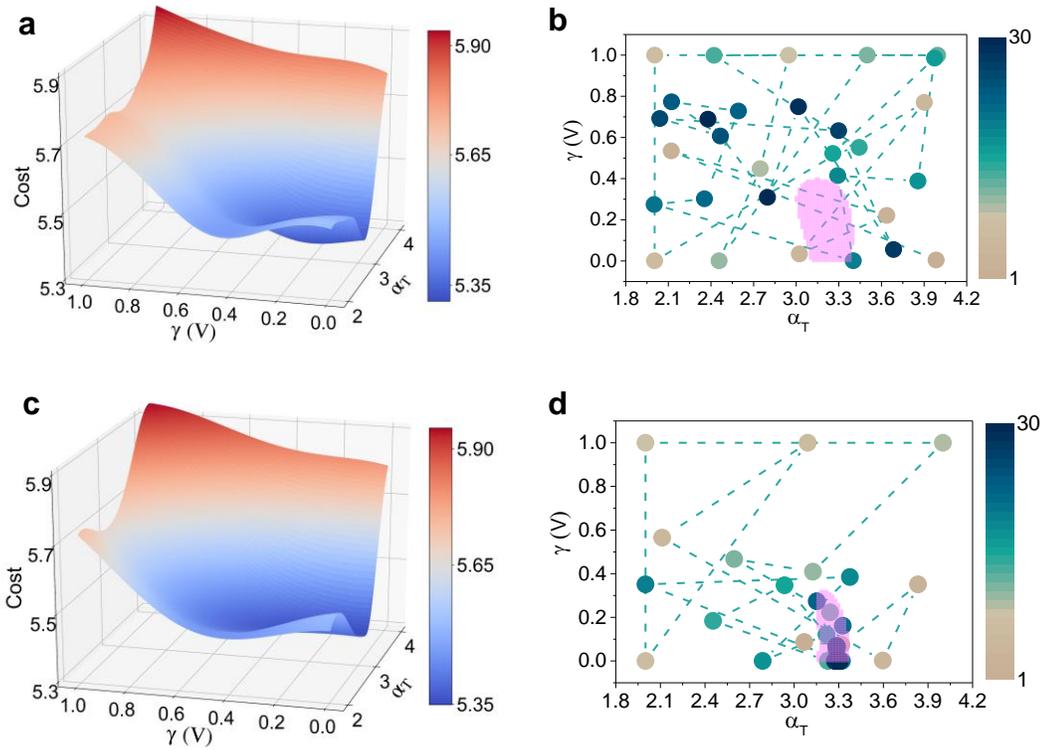

Figure S5. MCMC-BO simulation result with acquisition function (a)(b) PI, and (c)(d) UCB. The definition of the parameters and variables are the same as that in the main text.

The BO in the main text uses an EI acquisition function. To examine the effect of the acquisition function on the optimization results, results of two other forms of the acquisition functions in the forms of POI and UCB are obtained here, as shown in Figure S5. It is shown that the results are not sensitive to the specific choice of the acquisition function for the design optimization problem explored here. In addition, Bayesian optimization allows the exploration and exploitation of an optimum solution to be balanced by the choice of a model parameter. In exploitation, the design space near the minimum mean value is preferred, whereas, in exploration, the design space with large uncertainty is preferred. In the EI acquisition function, the margin parameter $\delta$, controls the tradeoff, in which a smaller value of $\delta$ prefers exploitation whereas a larger value prefers exploration[S4].

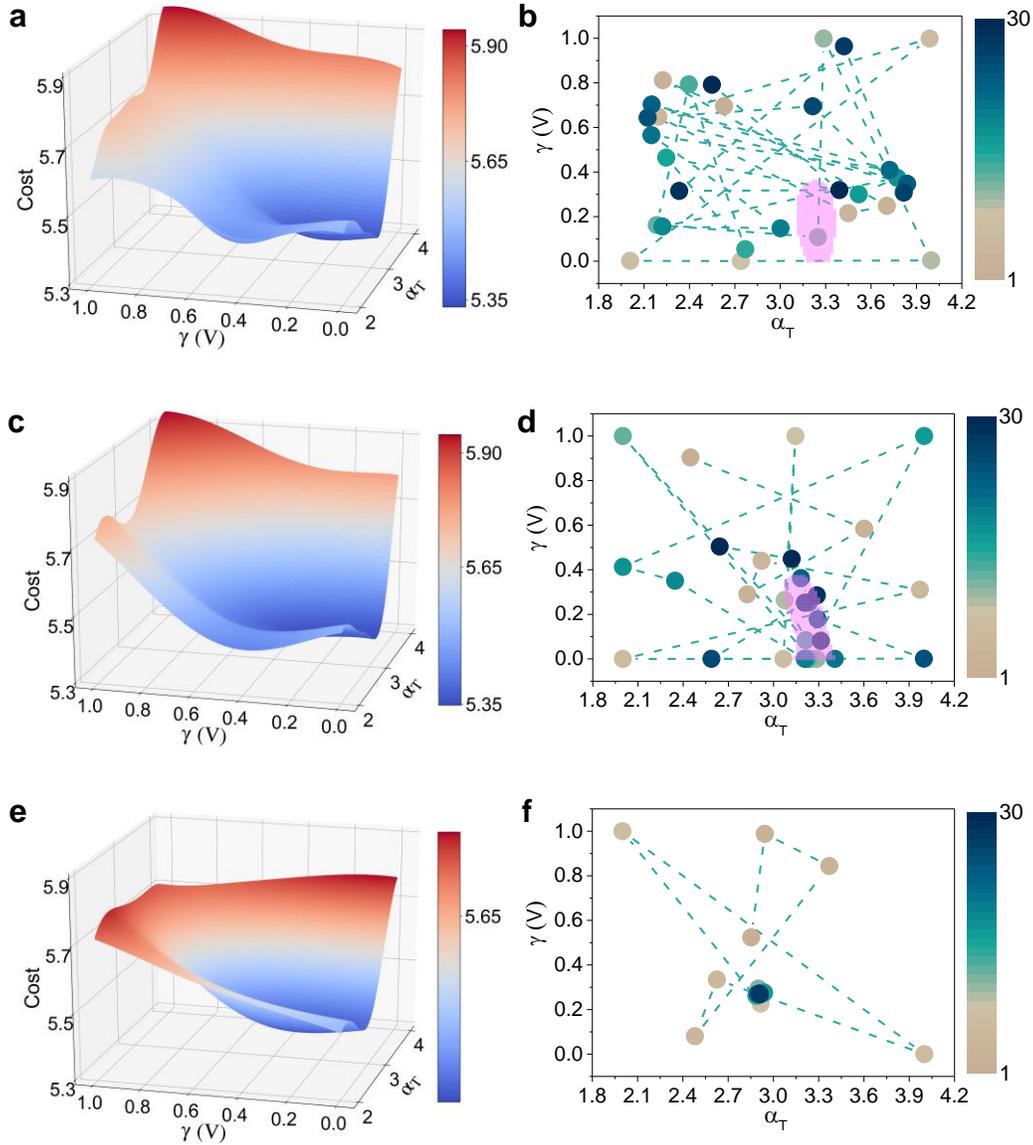

Figure S6. Exploitation vs. exploration in BO design optimization using EI method with (a)(b) $\delta = 3$, (c)(d) $\delta = 0$, and (e)(f) $\delta = -3$. In (f), multiple sample points overlap in the dash-circled region. The definition of the parameters and variables are the same as that in the main text.

Figure S6 shows the results from the BO optimization with the EI acquisition function and different values of $\delta$ after 30 iteration steps. Preference of exploration results in sample points spreading out in the design space, as shown in Figure S6a. In comparison, the preference of exploitation results in focused sampling near a small region, as shown in Figure S6c. Different balance between exploration and exploitation results in different traces in the BO iteration steps, but a similar optimum region is identified for the design problem studied here.

**S10.    Effect of the stochastic crossbar array device**

In the discussions in the main text, the variations of the crossbar array (CBA) parameters are neglected. Figure S7 explores the effect of the randomness of the CBA parameters as another design parameter. The normalized conductance values in the crossbar array are assumed to obey the same Gaussian distribution. For each individual conductance value normalized to its mean value, the Gaussian distribution has a standard deviation of $\xi_{CB}$.

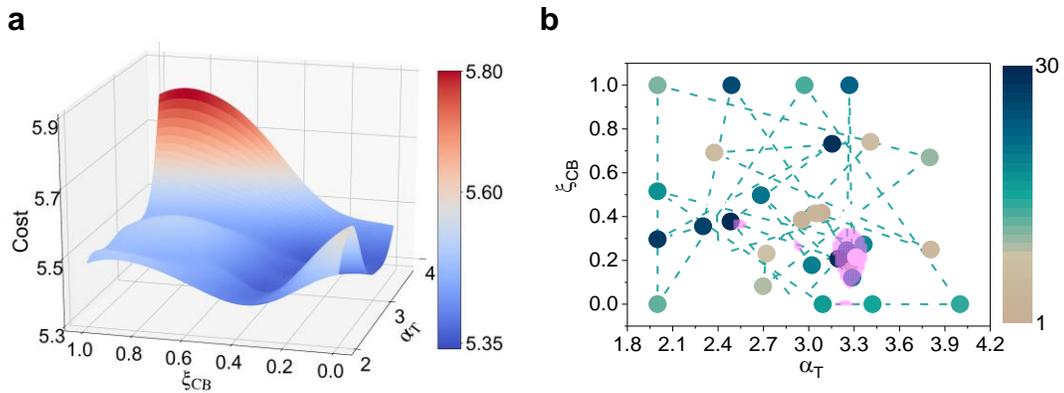

Figure S7. Design with variations of the crossbar array parameter in simulated annealing. The acquisition function is EI method. The definition of the other parameters and variables are the same as that in the main text.

Figure S7 identifies the optimum region in the ($\xi_{CB}$, $\alpha_T$) space, which forms a valley in the design space. To maintain the similar optimum BM performance, a decrease in the randomness of the CBA can be compensated by an increase in $\alpha_T$ within the optimum region.

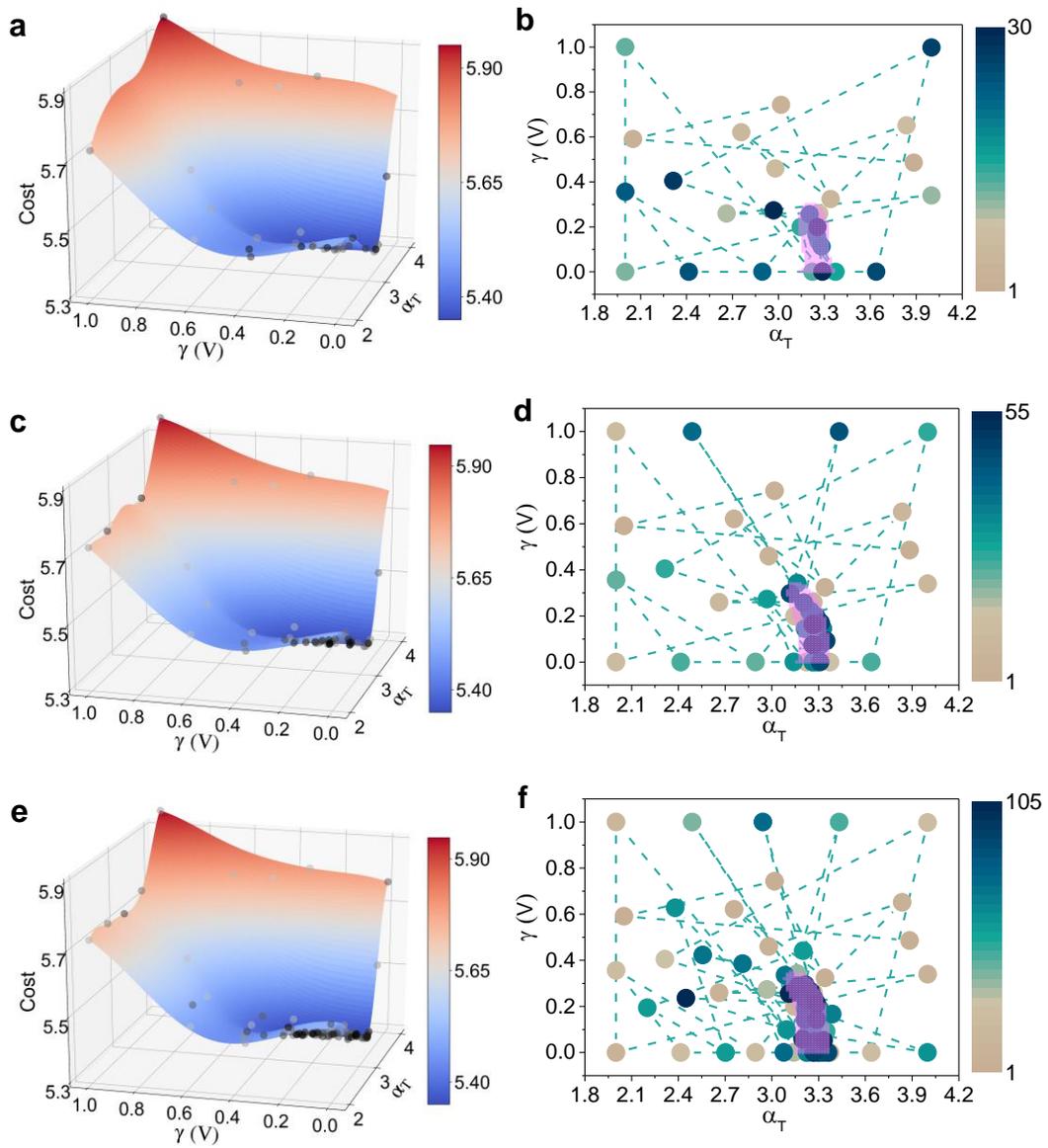

Figure S8. MCMC-BO design simulations for the stochastic parameters of the Boltzmann machine with different numbers of iteration steps, (a)(b) 25, (c)(d) 50, and (e)(f) 100 steps. The definition of the parameters and variables are the same as that in the main text.

## S11. Number of iteration steps in the BO

Figure S8 compares the typical results after different numbers of BO iteration steps in one simulation. Comparison of the optimum region identified after 25, 50, and 100 BO iteration steps indicates that even with 25 BO steps, the identified optimum is close to that with 100 BO steps. To quantify the relative difference between the identified optimum after $n$ BO steps to that of *m* steps, we compute the relative error $e_o(n,m) = |o_{\min}^n - o_{\min}^m|/o_{\min}^m$, where $o_{\min}^n$ is the minimum of the average cost function after *n* steps. The relative difference between 25 and 50 steps is 0.011%, and that between 25 and 100 steps is 0.026%. The results confirm that the 25-iteration MCMC-BO method is a good balance between the computational cost and accuracy.